\documentclass{elsarticle}
\usepackage[modulo]{lineno}
\usepackage{amsmath, amsthm, amssymb, amsfonts} % pour les symboles uft8
\usepackage{amssymb} % pour les symboles uft8
\usepackage{wasysym} % many more characters
\usepackage[sicmds,freestanding]{hepunits}
\robustify\dots
\usepackage{cleveref}
\usepackage{subfig}
\usepackage{ upgreek }
\usepackage{graphicx}

\usepackage{newunicodechar}
\newunicodechar{σ}{\ensuremath{\sigma}}

\begin{document}
% \sffamily

\title{Optimisation of a silicon-tungsten electromagnetic calorimeter energy response to photons}
\author[1]{Y. Shi\fnref{fn1}}
\ead{shi@llr.in2p3.fr}
\author[1]{V. Boudry\corref{cor1}}
\ead{boudry@llr.in2p3.fr}
\fntext[fn1]{This is the first author footnote}
\cortext[cor1]{Corresponding author}
\affiliation[1]{organization={Laboratoire Leprince-Ringuet, CNRS, École polytechnique, Institut Polytechnique de Paris,},
                %addressline={Street 29}, 
                postcode={91128}, 
                %postcodesep={}, 
                city={Palaiseau},
                country={France}}

\begin{abstract}
An innovative path for the detectors at future colliders to achieve higher performances is to use a Particle Flow approach, which requires highly granular calorimeters to image individual showers. The silicon–tungsten electromagnetic calorimeter (SiW-ECAL) aims at fulfilling all the expected physical and technical requirements. %%
 SiW-ECAL has been developed by the CALICE and ILD collaborations for more than two decades and is now reaching maturity, for linear machines. %%
However, with the tendency towards circular machines, the progress of electronics and the rapid advancement of machine learning (ML) techniques, the SiW-ECAL design needs to be reoptimised to enhance its performance.
% especially in the low-energy range and to fully exploit ML-based energy reconstruction methods. %%
This study develops ML-based reconstruction approaches for SiW-ECAL, achieving an approximate \qty{20}{\%} improvement in energy resolution in the low-energy range and effectively correcting energy leakage in the high-energy range. Subsequently, the SiW-ECAL design is re-optimized based on this method.

\end{abstract}

\maketitle

\section{Introduction}

A key performance for experiments at future e$^+$e$^-$ Higgs, Electro-Weak and Top (HET) factories, such as FCC-ee, CEPC, ILC, CLIC, is the jet energy resolution. A typical example is the separation of the ZZ and WW final states, which requires a Jet Energy Resolution(JER) of \qty{3.8}{\%}, or better\,\cite{bartmann_future_2025, group_cepc_2025, abramowicz_ild_2025}. 

The use of Particle Flow Algorithms (PFA)\,\cite{brient_calorimetry_2002,brient_particle_2009} is an effective approach to achieve this goal, by reconstructing all final-state particles individually, using the optimal sub-detector for each type. Charged particles are generally best measured with the trackers, in which case the calorimeters are used only for neutral particles. The PFA requires a high-precision tracking and high granularity calorimeters, to topologically separate contributions of particle showers, using a centimetric segmentation~\cite{thomson_particle_2009}. The PFA is particularly well suited for the HET factories, where the produced bosons predominantly decay in multi-jets of medium energy (\qtyrange{45}{120}{GeV}).

High-granularity calorimeters have been studied by the CALICE collaboration for many years, with the silicon–tungsten electromagnetic calorimeter (SiW-ECAL) being one of them. The SiW-ECAL was initially designed for experiments at the ILC\,\cite{behnke_international_2013, the_ild_collaboration_international_2020}; however, it has become increasingly necessary to reoptimise its design, driven by the following three factors:

\begin{enumerate}
\item The particles in a jet are concentrated in the low-energy region; the neutral and low-energy particles exhibit particular behaviours w.r.t. their high-energy counterparts and require adequate treatments; %
e.g. it has been known for some time\,\cite{reinhard_calice_2009} that for photons in highly granular calorimeters, such as the SiW-ECAL, the energy resolution obtained by counting hits (cells above a given threshold) outperforms that obtained by summing the energies of the hits when the particle energy is below $\sim\qty{2}{\GeV}$, which accounts for approximately \qty{90}{\%} of photon energy in a jet, but this has somehow never been fully integrated in the optimisation.

\item The SiW-ECAL was previously optimized based on PFA's\,\cite{thomson_particle_2009} and some specific ($\tau$'s) final states\,\cite{tran_reconstruction_2016}. Recently, a study on B mesons decays at Z pole indicated the importance of high granularity and energy resolution for {$\pi^0$}'s reconstruction\cite{Brient:2026akq}. With the rapid advancement of machine learning (ML) techniques, energy measurements can now benefit significantly from ML models such as GNN, PointNet, DGCNN, ParticleNet and so on\cite{scarselli2008GNN, qi2017pointnet,phan2018dgcnn,qu2020particleNet}. It is therefore natural to reoptimise the SiW-ECAL to take full advantage of ML methods.

\item Circular colliders are proposed for future Higgs factories, the maximum centre of mass energy of them is lower than for linear colliders, which requires less material budget of ECAL.
\end{enumerate}

This study develops an ML-based energy reconstruction method, with an emphasis on the low-energy region, and reoptimised the SiW-ECAL with this method. This article is organized as follows: \cref{sec:linearity_and_resolution} introduces the basic chain to derive calorimeter performance. \Cref{sec:ml_method} presents the ML-based energy reconstruction approach. \Cref{sec:reoptimization} discusses the reoptimisation of the SiW-ECAL, and \cref{sec:conclusion} provides the conclusions and outlook.

\section{Energy linearity and resolution}
\label{sec:linearity_and_resolution}

Energy linearity and resolution are the key performance indicators of calori\-meters. In this study, the SiW-ECAL performance was obtained through the basic chain consisting of three procedures: simulation, reconstruction, and evaluation.

\begin{figure}
    \centering
    \includegraphics[width=0.45\linewidth]{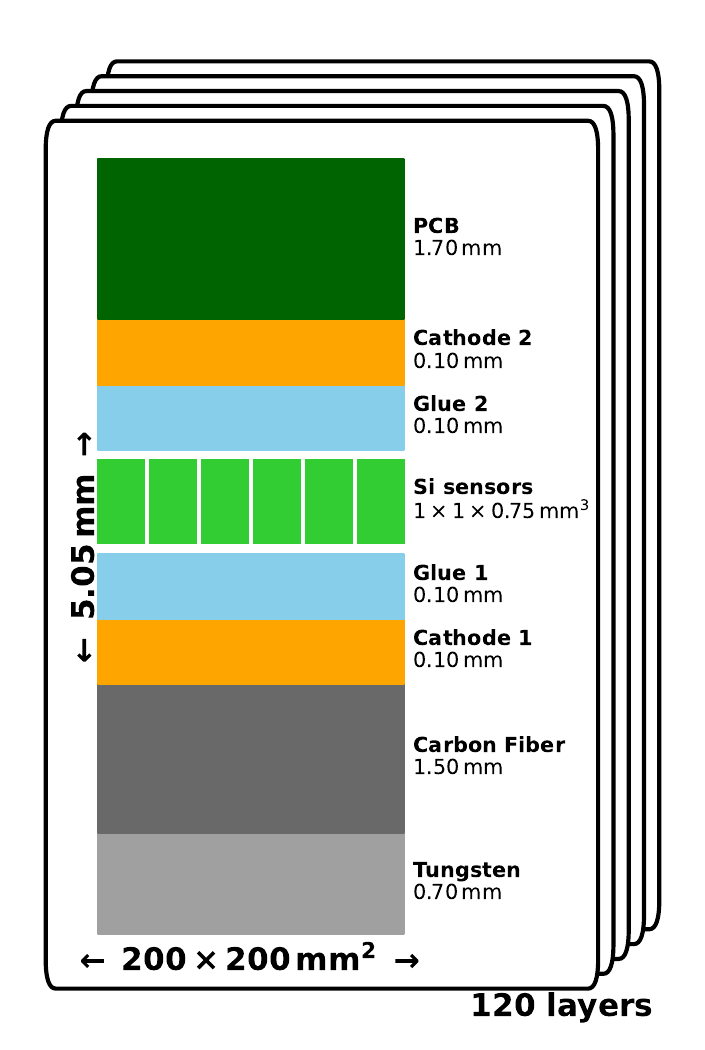}
    \caption{Schematic diagram of the SiW-ECAL prototype geometry.}
    \label{fig:ECALGeo}
\end{figure}

The simulation was performed using DD4hep\,\cite{frank_markus_2018_1464634}, a software framework built with ROOT and Geant4. A flexible SiW-ECAL prototype geometry was implemented, as illustrated in \cref{fig:ECALGeo}. It consists of 120 layers, each comprising a tungsten absorber of \qty{0.7}{\mm}, approximately \qty{0.2}{X_{0}} in radiation length, and the silicon of \qty{0.75}{mm}. Including carbon fibre, cathodes, glue, and PCB, the total material of the prototype reaches about \qty{26}{X_0}. In the transverse direction, the prototype measures \qtyproduct{200x200}{\mm} — approximately \qty{2}{R_{M}}—which is sufficient to contain electromagnetic (EM) showers. The silicon is finely segmented into \qtyproduct{1x1}{\mm} cells. Such an ultra-segmented geometry is designed to provide a convenient way to simulate less-segmented ECAL geometries by selecting readout layers and combining Si cells without rerunning the simulation. For instance, a geometry with 30 sampling layers and \qtyproduct{5x5}{\mm} Si cells can be obtained by reading 1 in each 4 layers and combining 25 cells into 1. For reference, this 30-layer, \qtyproduct{5x5}{\mm} configuration is used as the baseline geometry in this study, while alternative geometries are discussed in \cref{sec:reoptimization}.

The generation includes two kinds of photon datasets: one with continuous energies ranging from \qtyrange{0}{70}{GeV}, and the other with discrete energies of \qtylist[list-units = single]{0.1; 0.25; 0.5; 1.0; 2.0; 5.0; 10.0; 20.0; \dots; 60.0}{\GeV}. The continuous-energy data set contains 1.25 million events, of which 1 million are used for training and the rest for validation. The discrete-energy data set, with 50 thousand events per each energy point, is employed to evaluate the ECAL performances from different reconstruction methods or geometry configurations. It is referred to as the test sample, in contrast to the training sample and the validation sample. All photons are generated along the vertical direction without any angular spread. A threshold of 
\qty{0.1}{MIP} was applied to calorimeter hits, where \si{MIP} denotes the energy deposited by a minimum-ionizing particle traversing a Si cell.

The reconstruction employed different energy reconstruction methods. Traditional approaches, “Sum E” and “N Hits,” are introduced as baselines in this section, while ML-based methods are presented later in \cref{sec:ml_method}. The Sum E method reconstructs the particle energy from the total deposited energy, whereas the N Hits method relies on the number of hits. The distributions of deposited energy and number of hits both deviate from a Gaussian form at low energies. Several fitting functions were tested to describe this behavior, and the gamma function provides the best modeling—fitting well at low energies and converging to Gaussian at high energies. Therefore, the gamma fit is adopted in this study.

% \begin{figure}[htbp]
%     \centering
%     \subfloat[]{
%         \includegraphics[width=0.45\linewidth]{Figures/SumE_Fit.pdf}
%         \label{fig:SumE_Fit}
%     }
%     \hfill
%     \subfloat[]{
%         \includegraphics[width=0.45\linewidth]{Figures/Nhits_Fit.pdf}
%         \label{fig:Nhits_Fit}
%     }
%     \caption{Fits for \protect\subref{fig:SumE_Fit} Energy deposition and \protect\subref{fig:Nhits_Fit} Number of hits.}
%     \label{fig:Gamma_Fit}
% \end{figure}

Calibration was carried out using the validation sample. For the Sum E method, a linear fit was applied to the deposited energy as a function of the particle energy, while for the N Hits method both a linear fit and a spline fit were used. The resulting calibration curves were then applied to the test sample to reconstruct the particle energy. The reconstructed energy distributions were fitted with a gamma function to extract the energy linearity and resolution.

\begin{figure}[htbp]
    \centering
    \subfloat[]{
        \includegraphics[width=0.45\linewidth]{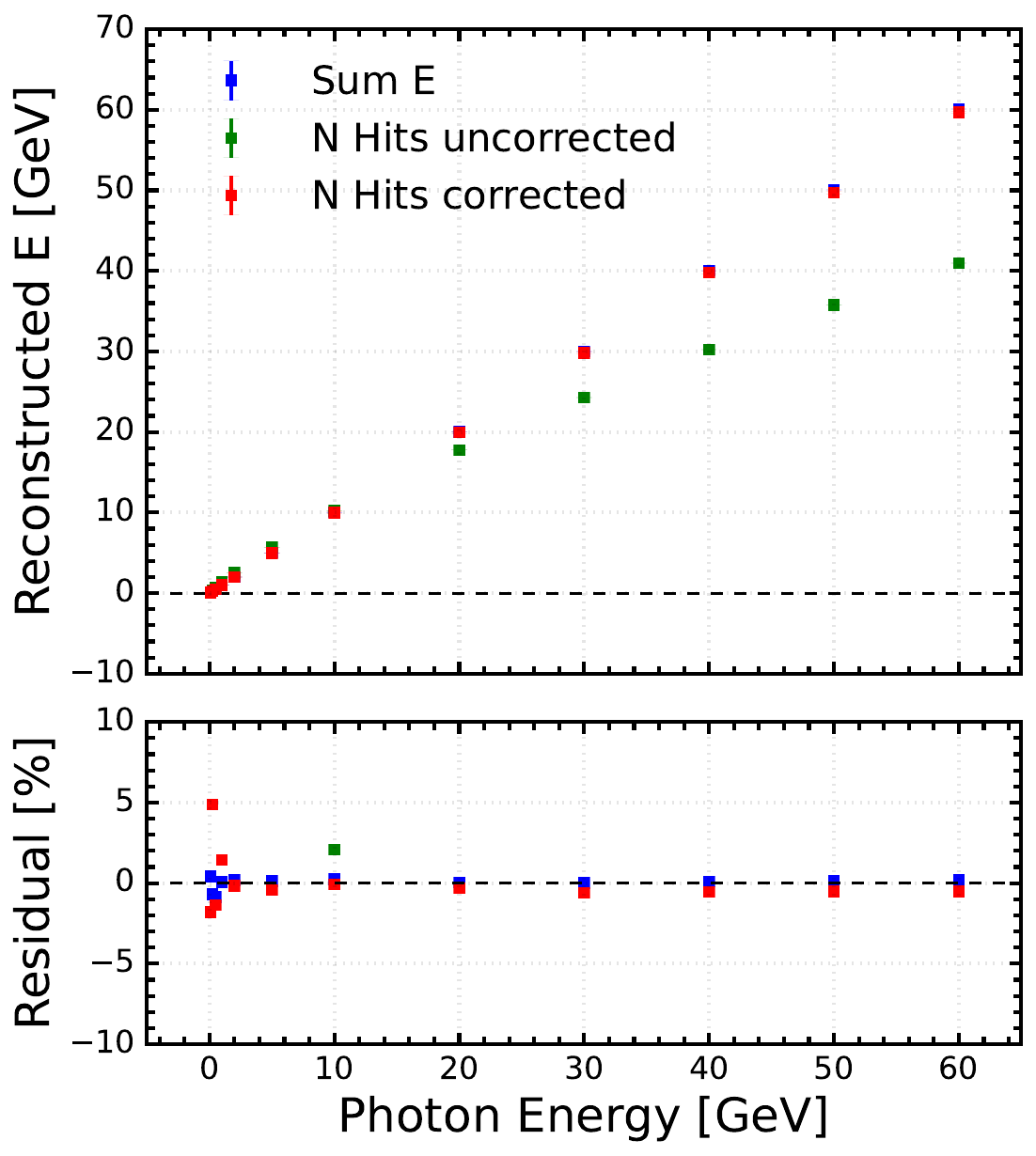}
        \label{fig:Linearity_basic}
    }
    \hfill
    \subfloat[]{
        \includegraphics[width=0.45\linewidth]{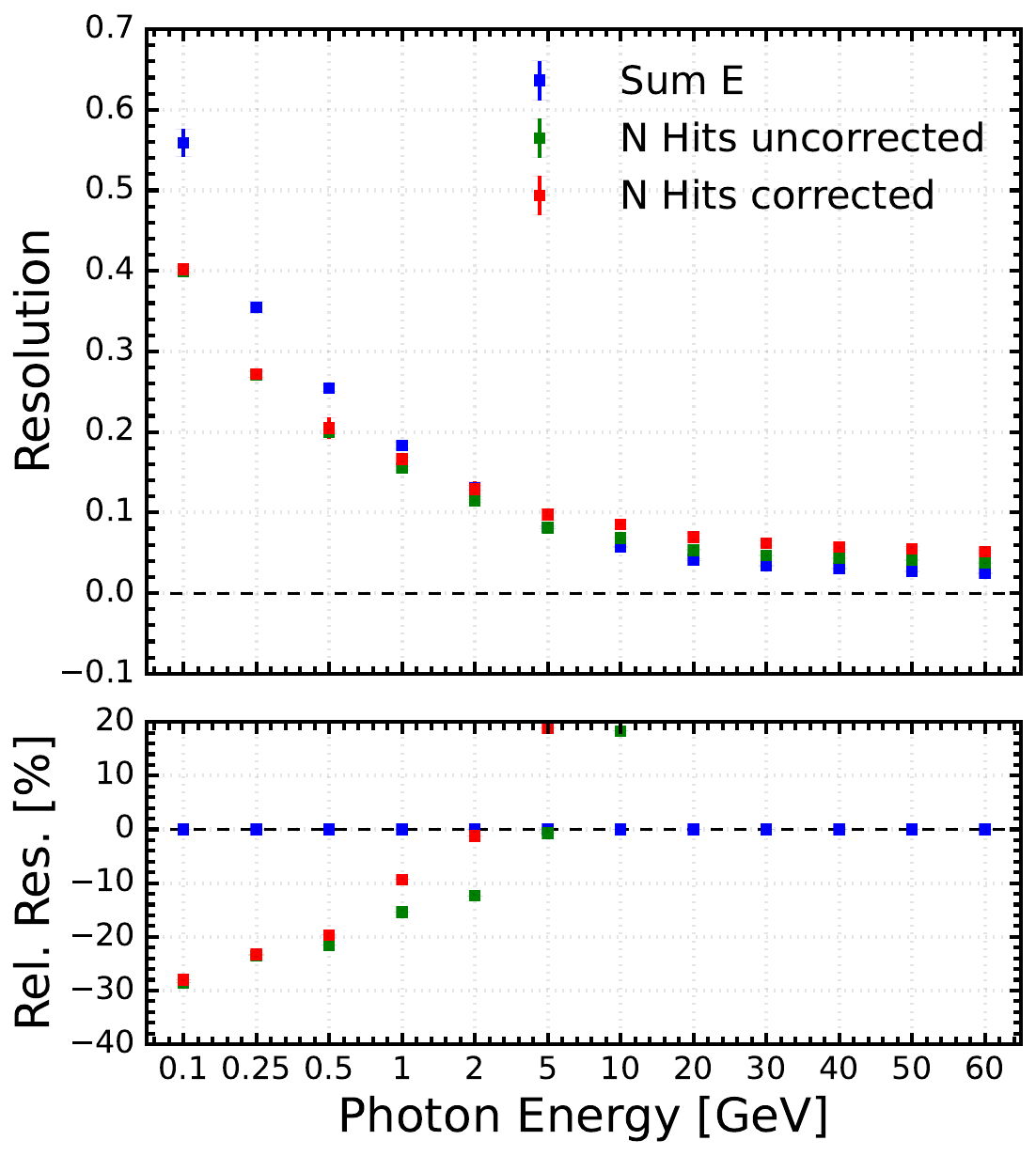}
        \label{fig:Resolution_basic}
    }
    \caption{\protect\subref{fig:Linearity_basic} Energy linearity and \protect\subref{fig:Resolution_basic} energy resolution for the Sum E and N Hits methods. The residual is defined as $(E_{\mathrm{reconstructed}} - E_{\mathrm{true}})/E_{\mathrm{true}}$ and "Rel. Res." denotes the Relative Resolution with respect to Sum E method.}
    \label{fig:Performance_basic}
\end{figure}

The obtained energy linearity and resolution are shown in \cref{fig:Performance_basic}. For the N Hits method, the uncorrected case corresponds to the linear fit calibration, whereas the corrected case corresponds to the spline fit calibration. The spline fit effectively corrects the non-linearity of the N Hits method at high energies, at the cost of a degraded resolution. The residual is defined as $(E_{\mathrm{reconstructed}} - E_{\mathrm{true}})/E_{\mathrm{true}}$, and the relative resolution (Rel. Res.) is defined with respect to the Sum E method for comparison.

Note that the reconstructed energy from the N Hits method saturates quickly at high energies, but provides better energy resolution than the Sum E method for energies below \qty{5}{GeV}. This is because, at low energies, the distribution of energy deposition gradually evolves into a Landau distribution, producing a right-hand tail that degrades the resolution. In contrast, the distribution of the number of hits is less affected by this effect, resulting in improved resolution. This result implies that the granularity of the SiW-ECAL provide additional information beyond just energy, suggesting that a method capable of exploiting this additional information could outperform both traditional methods.

\section{ML energy reconstruction}
\label{sec:ml_method}

In order to take advantage of the high granularity and to improve performance in the low-energy range, two ML models were used to develop energy reconstruction methods: the Multi-Layer Perceptron (MLP) and the Dynamic Graph Convolutional Neural Network (DGCNN). While the DGCNN is among the most powerful models for energy reconstruction in high-granularity calorimeters, it requires substantial computational resources for training and hyperparameter tuning, making it impractical for the ultimate goal of ECAL reoptimisation. The MLP, although the simplest ML model, can achieve comparable performance to the DGCNN with input features selected based on physical insight and proper fine-tuning. Its fast training speed and minimal hyperparameter requirements make it particularly well-suited for ECAL reoptimisation.

The baseline geometry was used throughout this chapter. For the baseline configuration, a total of 152 physical features were extracted:
\[
[E_i, N_i, E_i/E_\mathrm{sum}, N_i/N_\mathrm{sum}, E_i/N_i, E_\mathrm{sum}, N_\mathrm{sum}],
\]
where $E_i$ and $N_i$ stand for the energy sum and number of hits in the $i^\mathrm{th}$ layer, 
and $E_\mathrm{sum}$ and $N_\mathrm{sum}$ stand for the total energy sum and total number of hits in the calorimeter. These physics-motivated features were used as inputs to the MLP. Accordingly, the MLP architecture was configured with three hidden layers of sizes [256, 128, 64]

In this study, the loss function was found to have the greatest impact on ECAL performance. Three loss functions were tested: Mean Square Error (MSE), Relative Mean Square Error (RMSE), and the Huber loss function. Their definitions are:

\[
L_{\mathrm{MSE}} = \frac{1}{N}\sum_{i=1}^{N}\left(y_{\mathrm{pred}}^i - y_{\mathrm{true}}^i\right)^2
\]

\[
L_{\mathrm{RMSE}} = \frac{1}{N}\sum_{i=1}^{N}\left(\frac{y_{\mathrm{pred}}^i - y_{\mathrm{true}}^i}{y_{\mathrm{true}}^i}\right)^2
\]

\[
L_{\mathrm{Huber}} = \frac{1}{N} \sum_{i=1}^{N}
\begin{cases}
\frac{1}{2} r_i^2, & r_i < 0.05 \\
0.05 \times (r_i - 0.5 \times 0.05), & r_i \ge 0.05
\end{cases}
\]
where y denotes the photon energy, and \( r_i = \lvert (y_{\mathrm{pred}}^i - y_{\mathrm{true}}^i)/y_{\mathrm{true}}^i \rvert\) defines the relative error for the i-th event.

\begin{figure}[htbp]
    \centering
    \subfloat[]{
        \includegraphics[width=0.45\linewidth]{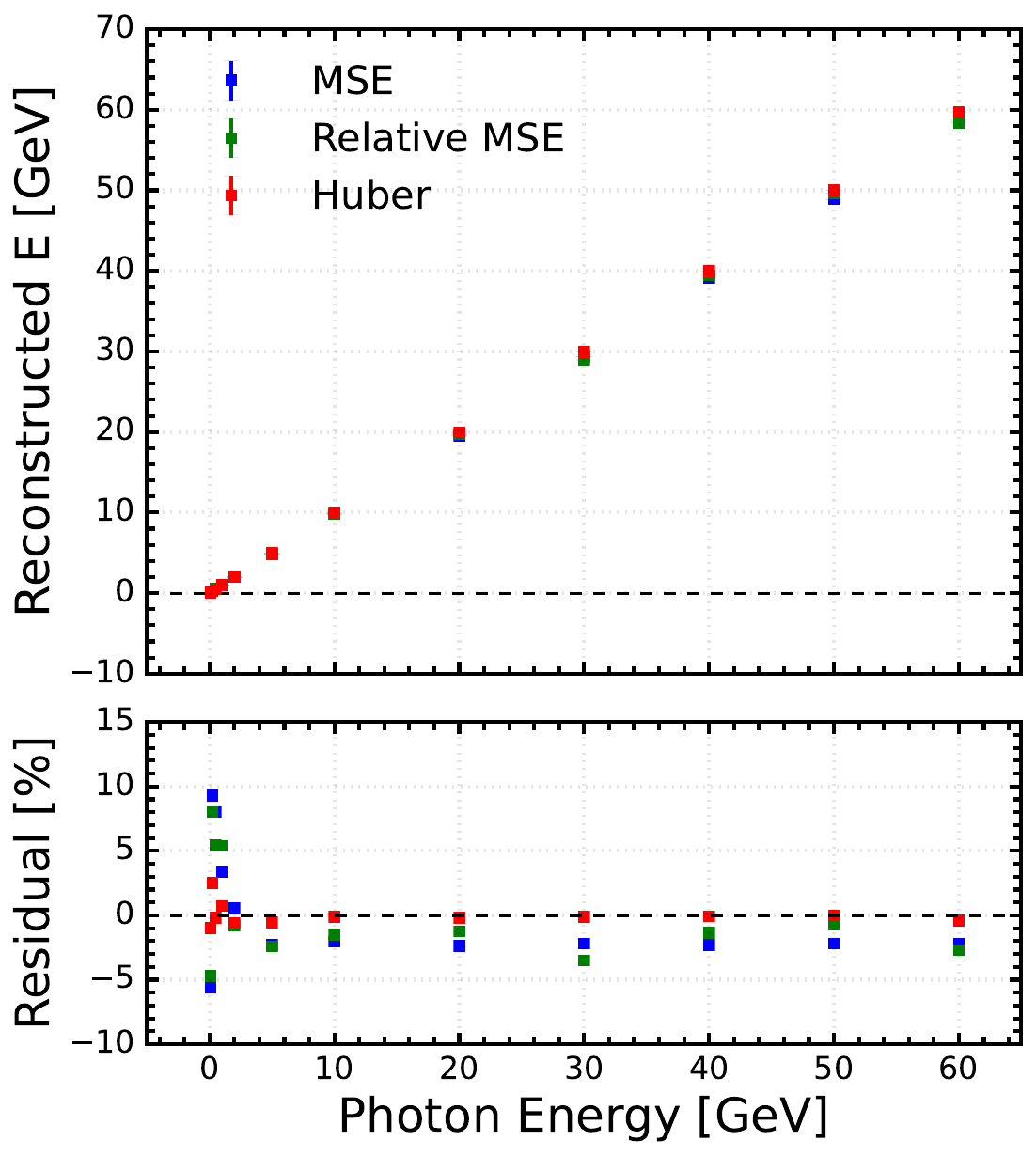}
        \label{fig:Linearity_huber}
    }
    \hfill
    \subfloat[]{
        \includegraphics[width=0.45\linewidth]{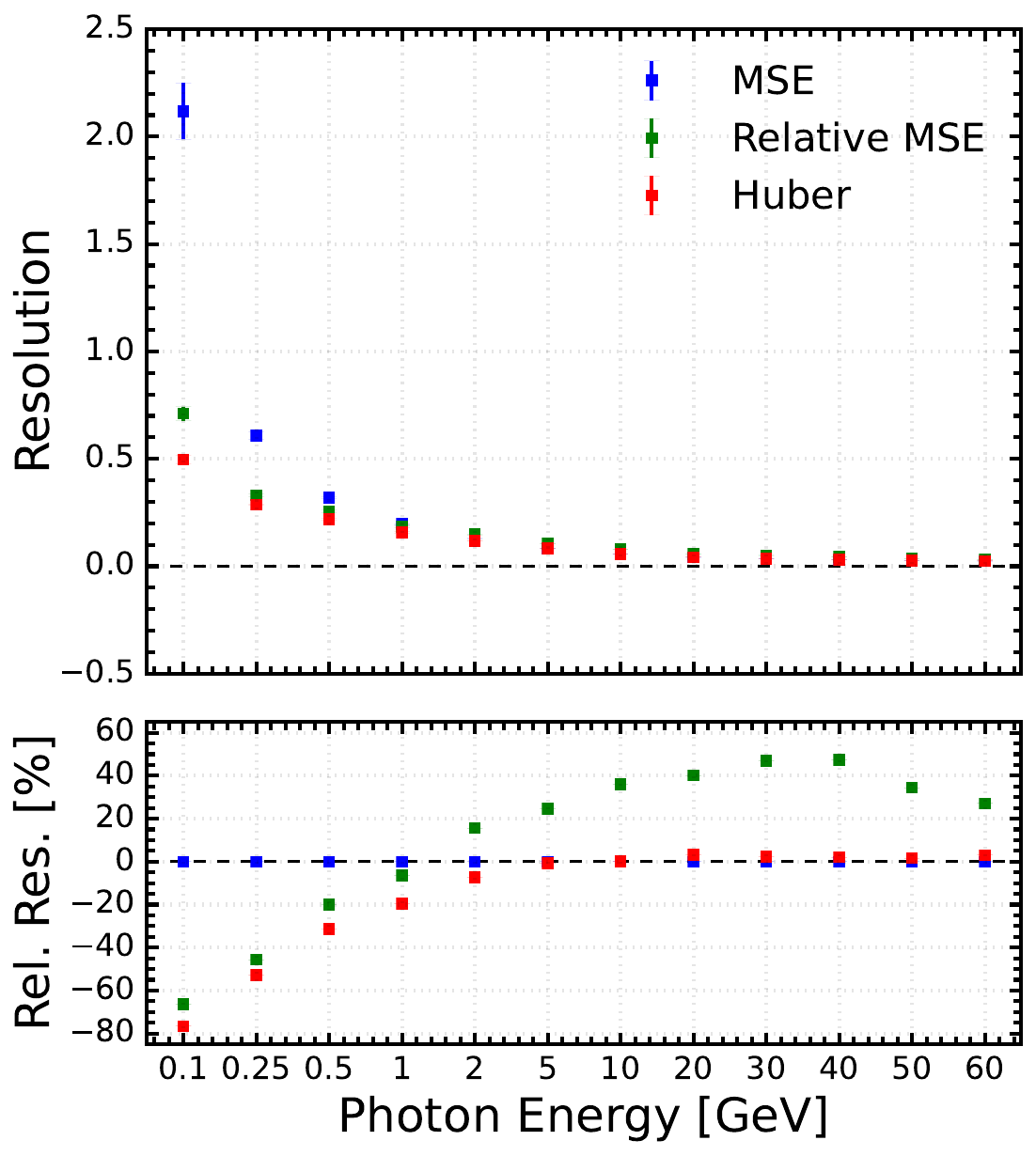}
        \label{fig:Resolution_huber}
    }
    \caption{MLP performance of different loss functions: (a)Energy linearity and (b)energy resolution (b).The residual is defined as $(E_{\mathrm{reconstructed}} - E_{\mathrm{true}})/E_{\mathrm{true}}$ and "Rel. Res." denotes the Relative Resolution with respect to the MSE.}
    \label{fig:Performance_huber}
\end{figure}

The performances of these three loss functions are shown in \cref{fig:Performance_huber}. Compared with $L_{\mathrm{MSE}}$, $L_{\mathrm{RMSE}}$ replaces the absolute difference with a relative difference, improving performance at low energies but deteriorating it at high energies. The $L_{\mathrm{Huber}}$ resolves this issue by introducing the condition $r_i < 0.05$, where $r_i$ represents the residual. Consequently, the ML model first learned to keep the linearity within 5\%, and then improved the resolution.

\begin{figure}
    \centering
    \includegraphics[width=0.45\linewidth]{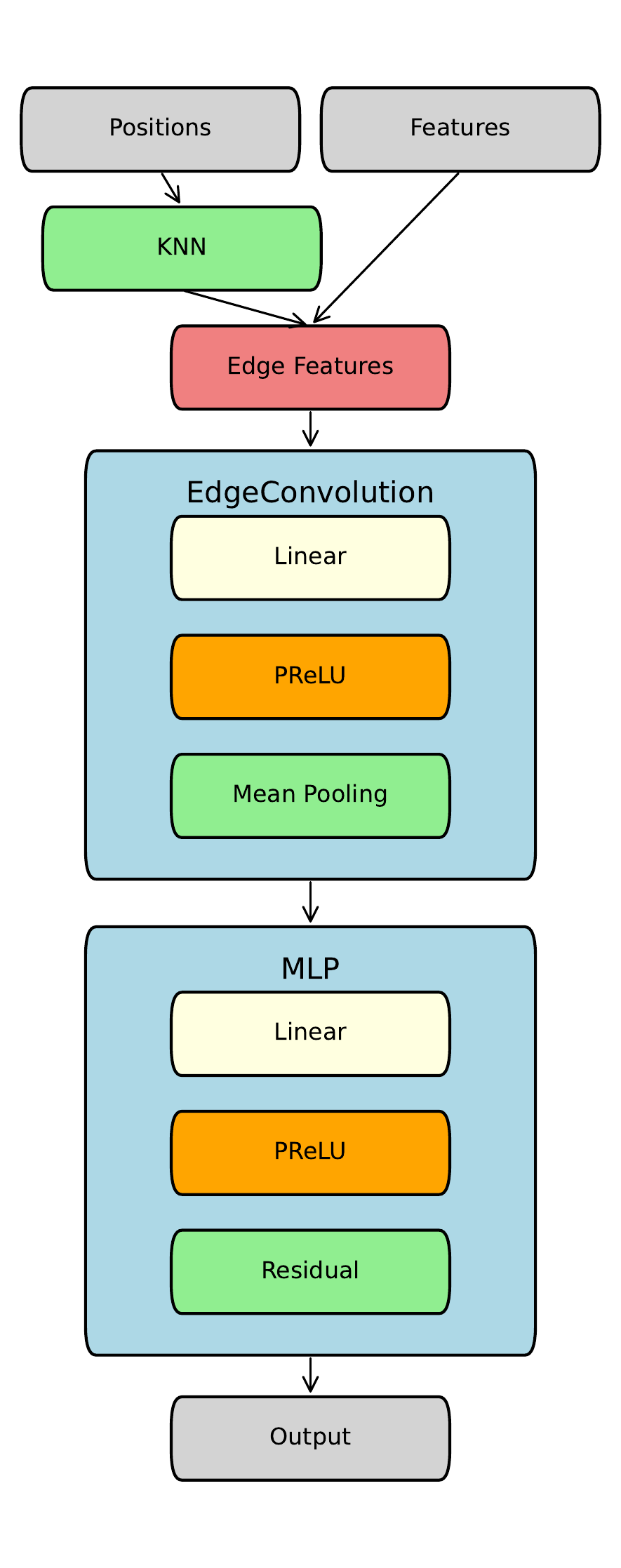}
    \caption{The architecture of the DGCNN model.}
    \label{fig:DGCNN_architecture}
\end{figure}

The DGCNN model is a state-of-the-art machine learning model for processing graph-structured data. The architecture of the model used in this study is illustrated in \cref{fig:DGCNN_architecture}. The key concept of this model is the edge convolution operation, where edges are constructed via the k-nearest neighbours (KNN) algorithm based on hit positions. Convolution extracts the local features encoded in the edges using small neural networks and the mean pooling method. These local features, together with the extra features, which are the features previously used in the MLP method, are then fed into the MLP module to predict the reconstructed energy.

\begin{figure}[htbp]
    \centering
    \subfloat[]{
        \includegraphics[width=0.45\linewidth]{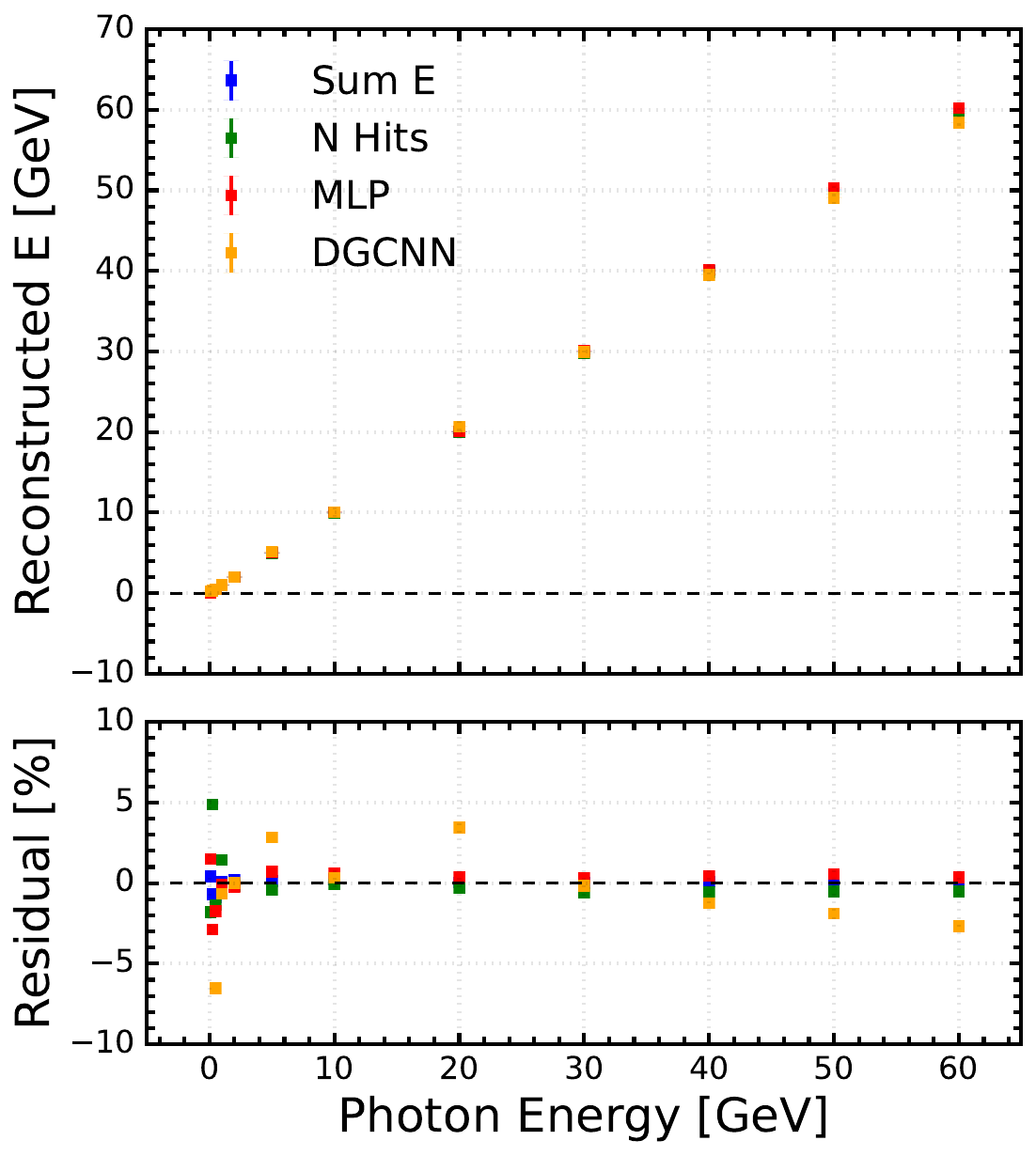}
        \label{fig:Linearity_ML}
    }
    \hfill
    \subfloat[]{
        \includegraphics[width=0.45\linewidth]{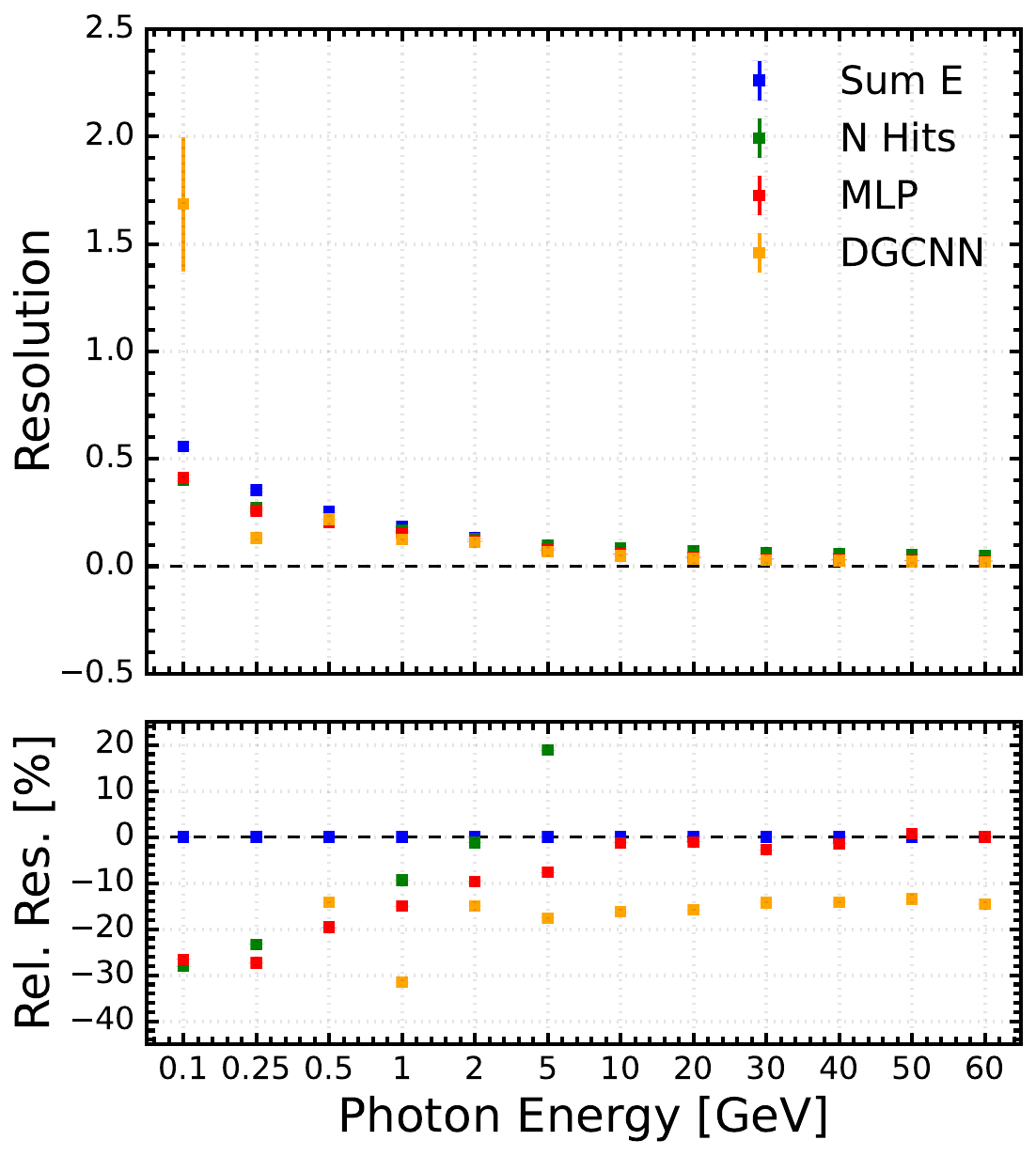}
        \label{fig:Resolution_ML}
    }
    \caption{ECAL performances of different energy reconstruction methods: (a)Energy linearity and (b)Energy resolution.The residual is defined as $(E_{\mathrm{reconstructed}} - E_{\mathrm{true}})/E_{\mathrm{true}}$ and "Rel. Res." denotes the Relative Resolution with respect to the Sum E.}
    \label{fig:Performance_ML}
\end{figure}

The performance of both ML-based and traditional methods is presented in \cref{fig:Performance_ML}. The MLP outperforms both the Sum E and N Hits approaches across the full energy range. While the DGCNN exceeds the MLP at high energies, its performance degrades at low energies, mainly because the KNN parameter k=32 causes nearly all hits connected to each other in low-energy events. Overall, the DGCNN does not outperform the MLP much but demands significantly greater computational resources, making it impractical for ECAL reoptimization. Therefore, the MLP is adopted for the ECAL reoptimization in the next section.
 
\section{ECAL reoptimisation}
\label{sec:reoptimization}

The ECAL reoptimisation was performed by comparing the performance of different calorimeter geometries. These geometries were derived from the baseline design shown in \cref{fig:ECALGeo} by selecting readout layers and merging Si cells. The MLP was used in energy reconstruction, with its architecture and hyperparameters optimized for each geometry. For example, when the ECAL sampling layers increased from 30 to 60, the hidden layer configuration in MLP was updated from [128,64,32] to [256,128,64]. 

\begin{figure}[htbp]
    \centering
    \subfloat[]{
        \includegraphics[width=0.45\linewidth]{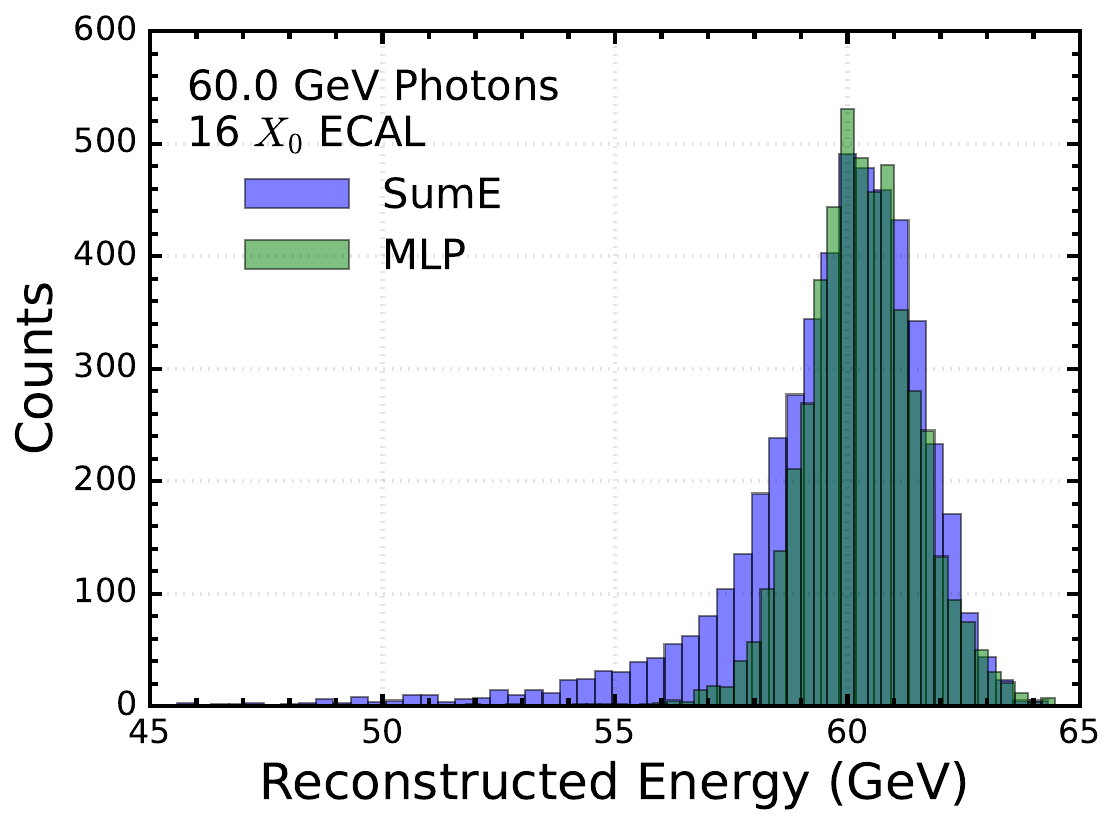}
        \label{fig:Hist_leakage_16X0}
    }
    \hfill
    \subfloat[]{
        \includegraphics[width=0.45\linewidth]{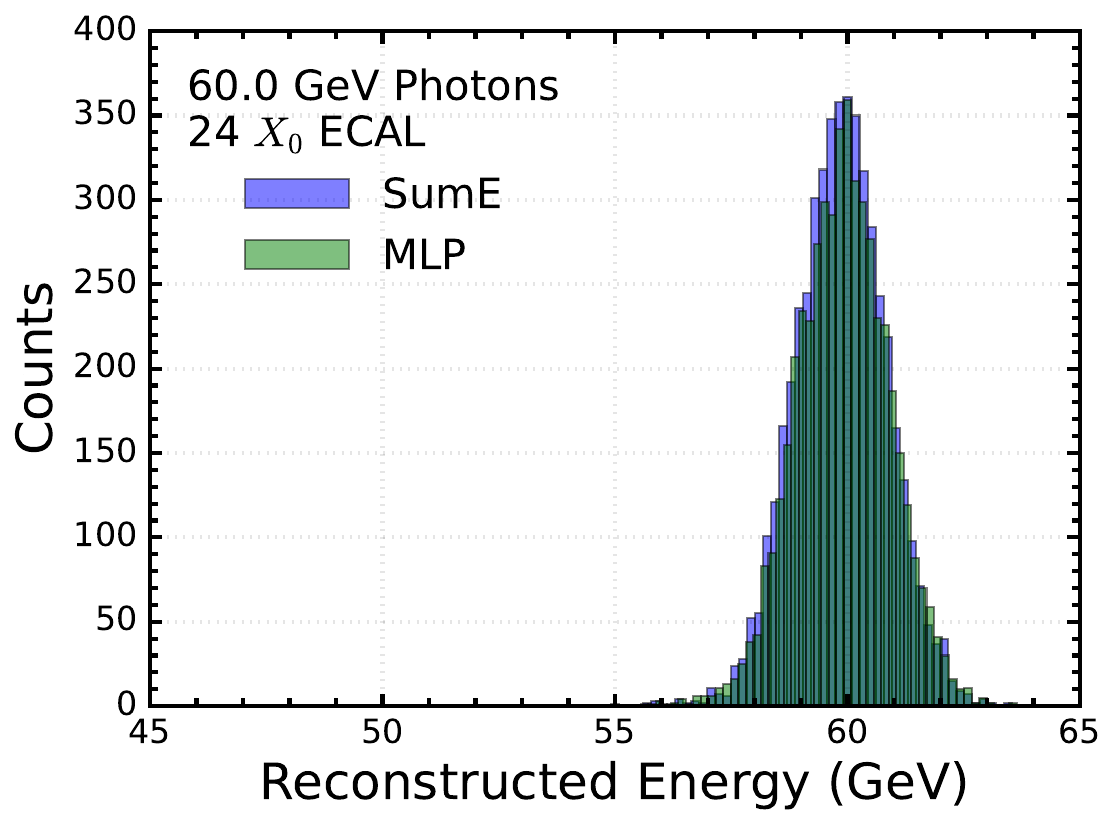}
        \label{fig:Hist_leakage_24X0}
    }

    \subfloat[]{
        \includegraphics[width=0.45\linewidth]{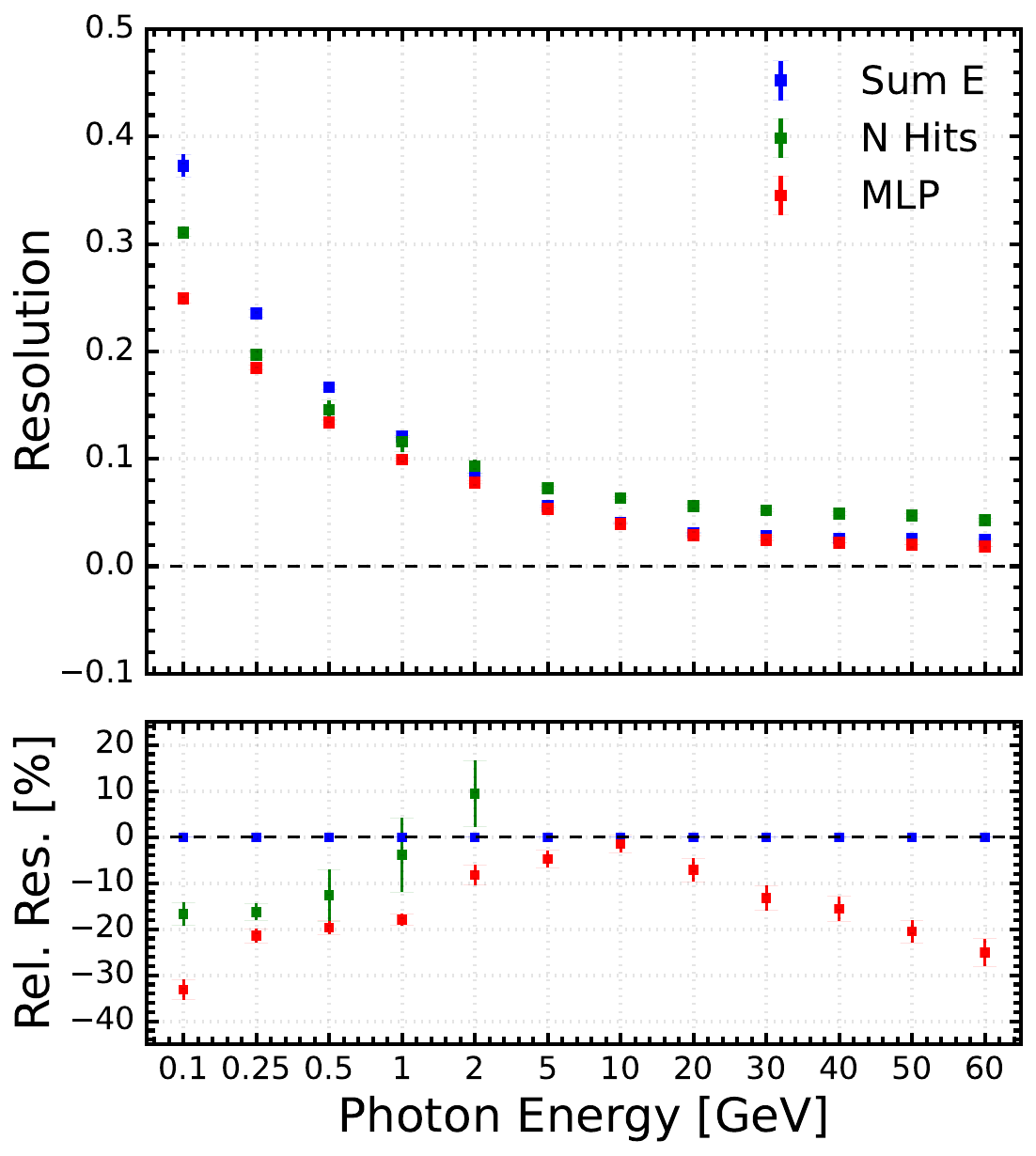}
        \label{fig:Resolution_leakage}
    }
    \hfill
    \subfloat[]{
        \includegraphics[width=0.45\linewidth]{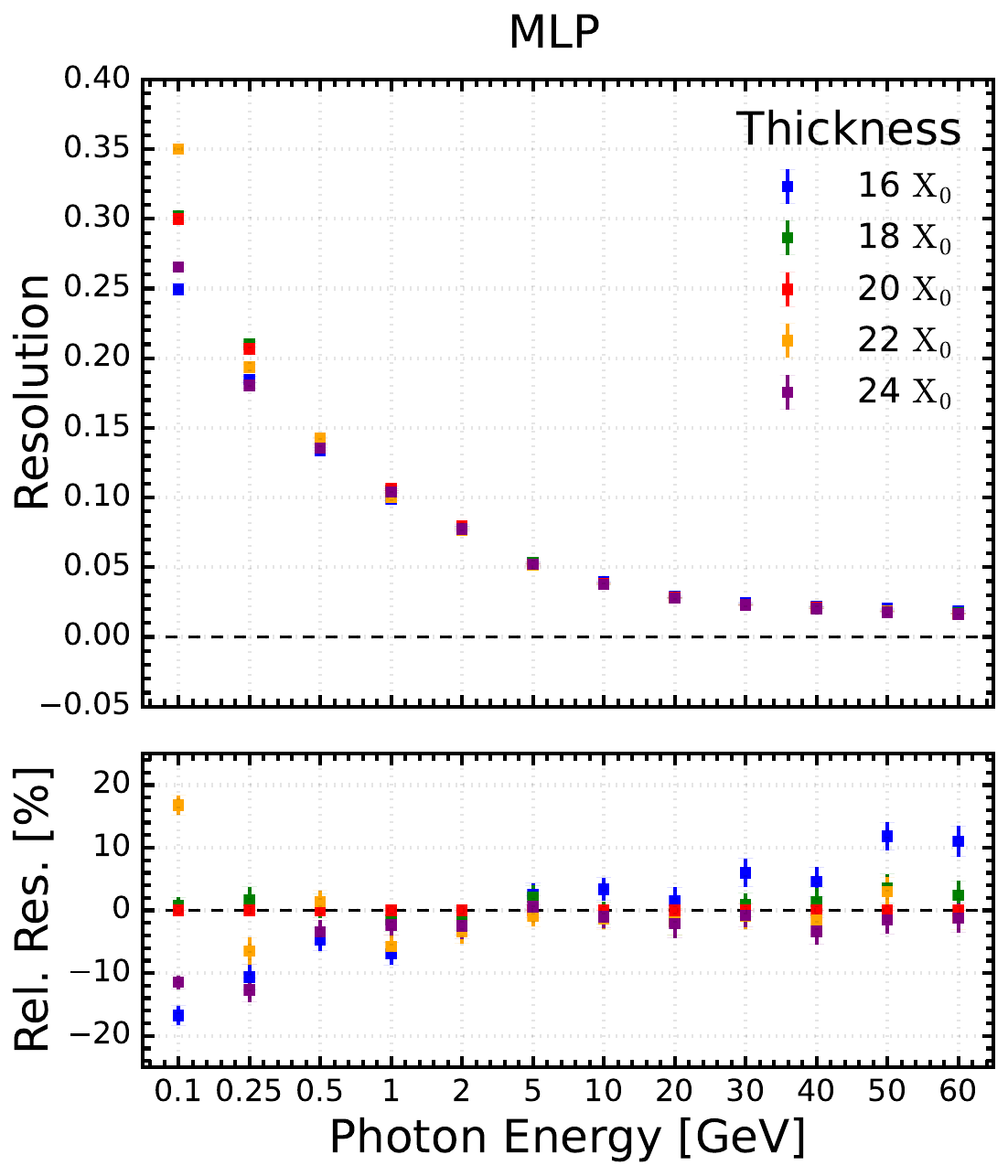}
        \label{fig:Resolution_X0}
    }
    \caption{
    Reconstructed energy distributions of \qty{60}{GeV} photons using the Sum E and MLP methods for \protect\subref{fig:Hist_leakage_16X0} \qty{16}{X_0} ECAL geometry and \protect\subref{fig:Hist_leakage_24X0} \qty{24}{X_0} ECAL geometry.
    \protect\subref{fig:Resolution_leakage} Energy resolutions of the \qty{16}{X_0} ECAL geometry reconstructed using the Sum E, N Hits, and MLP methods. 
    \protect\subref{fig:Resolution_X0} Energy resolutions of ECAL geometries with different absorber thicknesses from \qty{16}{X_0} to \qty{24}{X_0}. "Rel. Res." denotes the Relative Resolution.
}
\end{figure}

The ECAL radiation length was investigated by varying the number of readout layers used in reconstruction. Configurations with 80–120 layers were studied, corresponding to total absorber thicknesses of \qty{16}{X_0}–\qty{24}{X_0}, with the number of sampling layers adjusted accordingly to maintain a constant sampling ratio. Significant energy leakage was observed for the \qty{16}{X_0} configuration, as illustrated in \cref{fig:Hist_leakage_16X0}, where the Sum E method exhibits low-energy tails. Fortunately, this energy leakage could be largely corrected by the MLP, leading to a clear improvement in energy resolution, as shown in \cref{fig:Resolution_leakage}. However, even after MLP correction, the \qty{16}{X_0} configuration remains up to 10\% worse than the thicker tungsten configurations. It should also be noted that different geometries show noticeable differences at low energy. This arises because the loss function is dominated by high-energy data, which can introduce biases in the low-energy region. While larger models or careful tuning could mitigate this, the ECAL radiation length optimization focuses on high-energy performance. As a result, a minimum of \qty{18}{X_0} tungsten is recommended to suppress energy leakage.

\begin{figure}[htbp]
    \centering
    \subfloat[]{
        \includegraphics[width=0.45\linewidth]{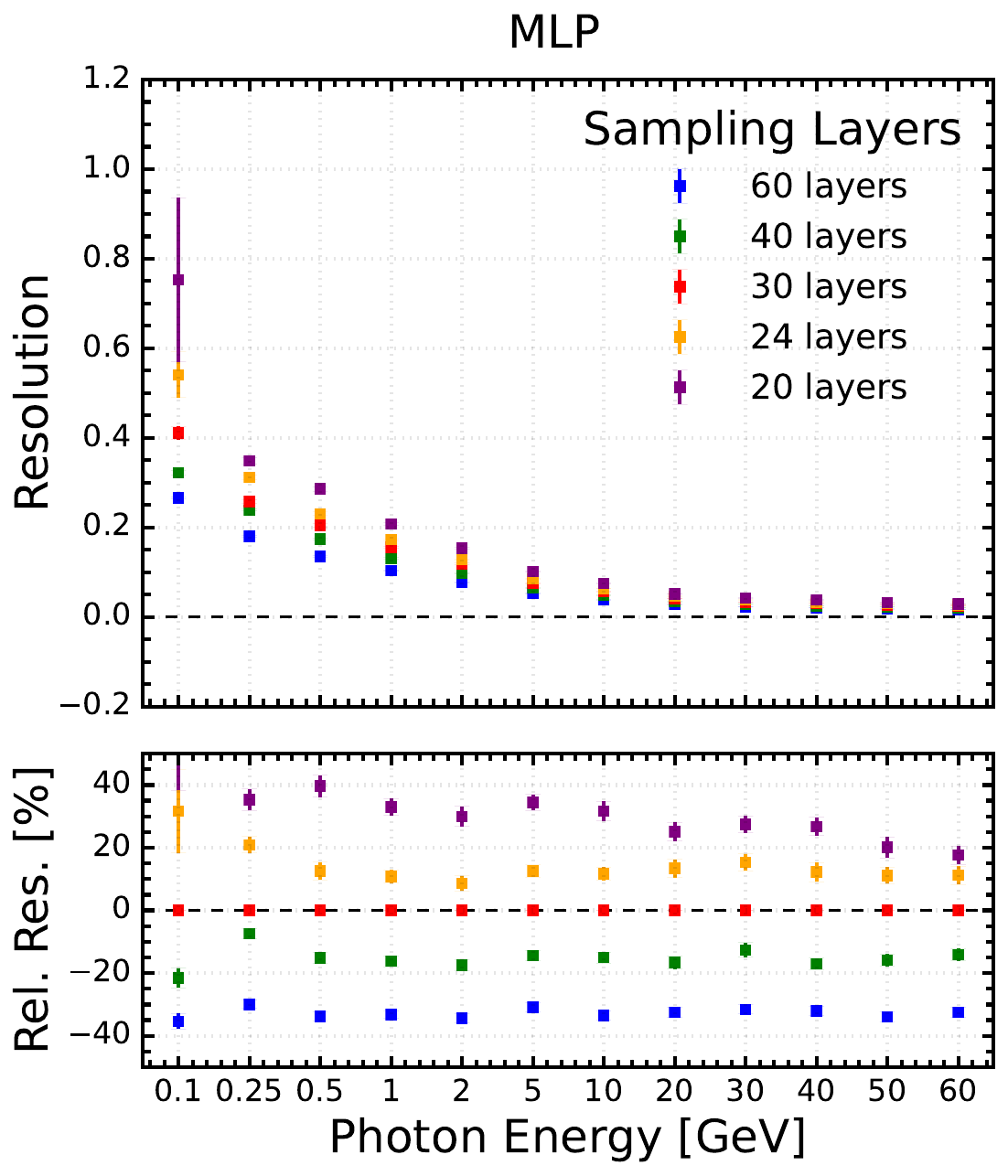}
        \label{fig:Resolution_SamplingLayer}
    }
    \hfill
    \subfloat[]{
        \includegraphics[width=0.45\linewidth]{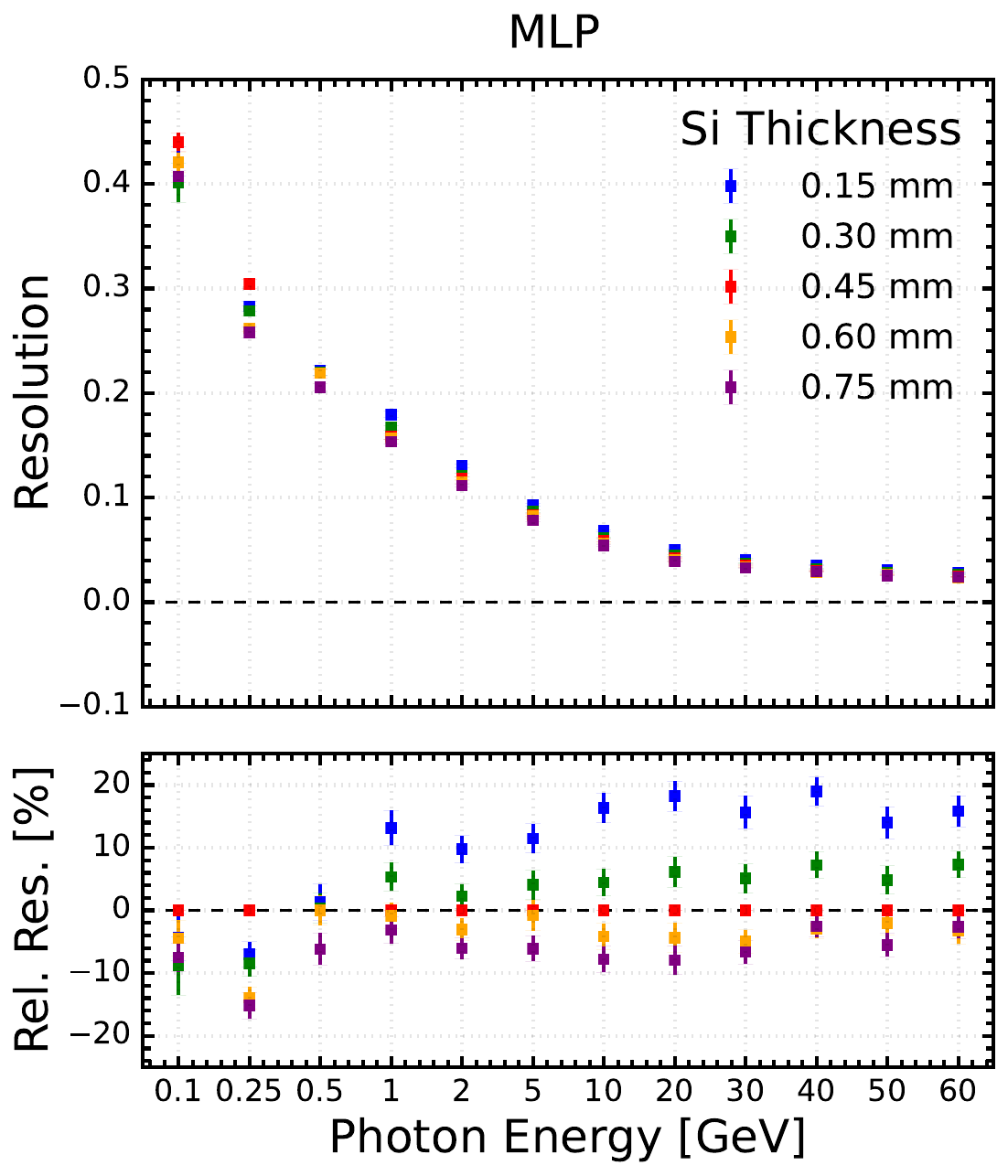}
        \label{fig:Resolution_SiThickness}
    }
\caption{
\protect\subref{fig:Resolution_SamplingLayer} Energy resolutions of ECAL geometries with different sampling layers.
\protect\subref{fig:Resolution_SiThickness} Energy resolutions of ECAL geometries with different Si thicknesses. "Rel. Res." denotes the Relative Resolution.}
\end{figure}

The number of ECAL sampling layers was investigated by selecting one readout layer every 2, 3, 4, 5, or 6 ECAL layers, corresponding to 60, 40, 30, 24, and 20 sampling layers. Their energy resolutions are shown in \cref{fig:Resolution_SamplingLayer}. The number of sampling layers has a significant influence on the energy resolution, particularly at low energies: when the number is reduced to 20, the limited number of hits makes it difficult for the ML model to converge. Although the performance consistently improves with an increasing number of sampling layers, the benefit saturates beyond 40 layers; therefore, 30 sampling layers are recommended.

The effect of the Si thickness is studied using a similar geometry, with the Si segmented longitudinally but fixed transversely with a cell size of \qtyproduct{5x5}{\mm}. ECAL configurations of different Si thickness were constructed by combining the segmented Si sub layers, energy resolutions of them are shown in \cref{fig:Resolution_SiThickness}. In general, thicker Si provides better energy resolution above \qty{1}{GeV}. With advances in semiconductor wafer production and manufacturing, \qty{0.75}{mm} Si is planned for the next version of the SiW-ECAL design.

\begin{figure}[htbp]
    \centering
    \subfloat[]{
    \raisebox{0.3\height}{
        \includegraphics[width=0.45\linewidth]{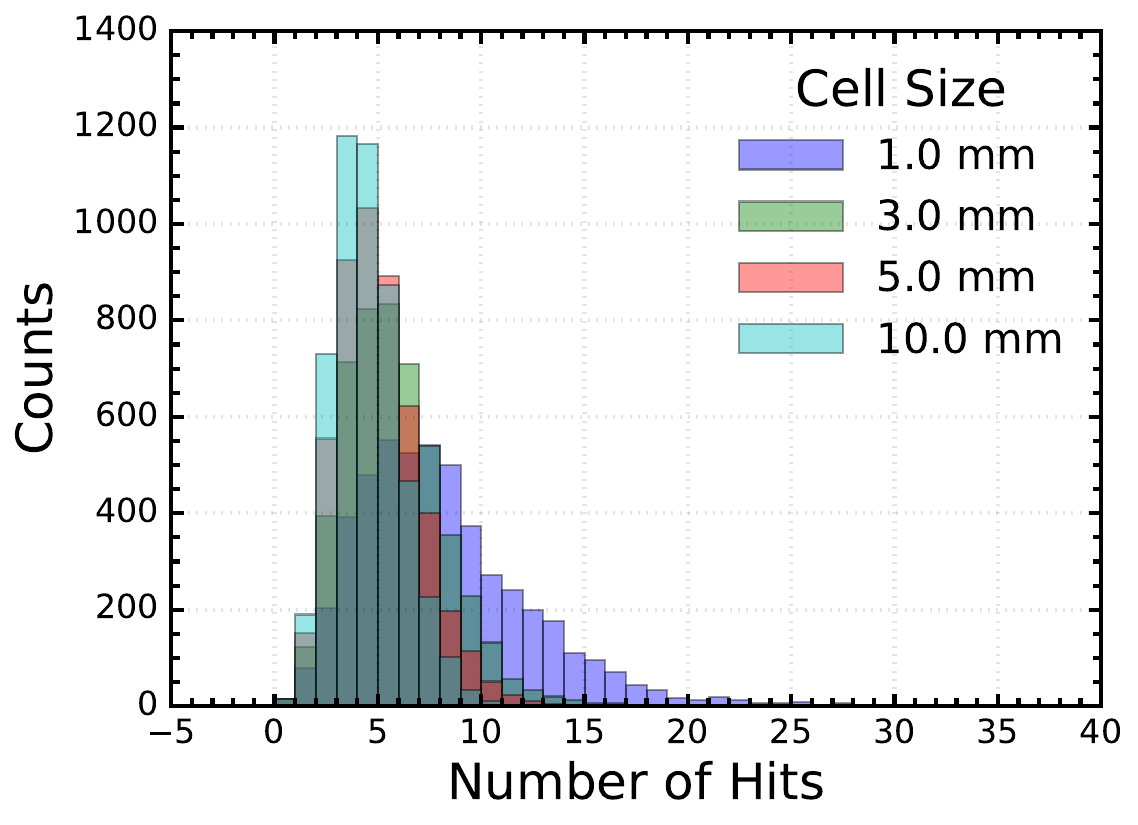}
    }
    \label{fig:Hist_cellsize}
}
    \hfill
    \subfloat[]{
        \includegraphics[width=0.45
        \linewidth]{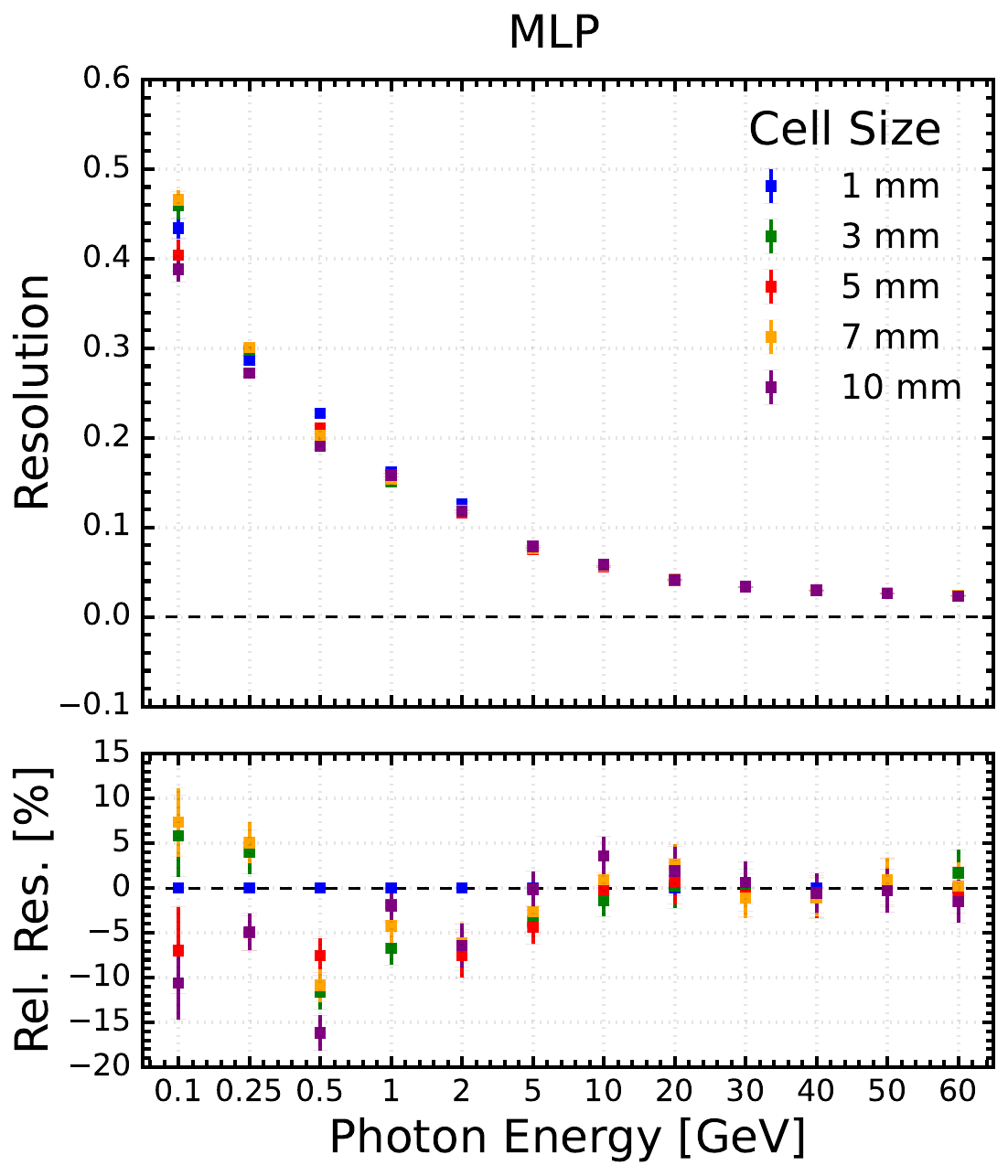}
        \label{fig:Resolution_SiCellSize}
    }
    \caption{
    \protect\subref{fig:Hist_cellsize} Number of hits distributions of ECAL geometries
    with different Si cell sizes for \qty{0.1}{GeV} photons.
    \protect\subref{fig:Resolution_SiCellSize} Energy resolutions of ECAL geometries with different Si cell sizes.  "Rel. Res." denotes the Relative Resolution with respect to \qty{1}{mm} cell size.
    }
\end{figure}

The Si cell size was evaluated by combining \qtyproduct{1x1x0.75}{\mm} cells into larger cells up to \qtyproduct{10x10x0.75}{\mm}. Counterintuitively, a smaller cell size does not improve the energy resolution. The underlying reason is that, when the cell size is smaller, a single particle is more likely to pass through multiple Si cells, producing a right-hand tail in the hit multiplicity distribution, as shown in \cref{fig:Hist_cellsize}, which ultimately degrades the energy resolution. 

The energy resolutions obtained for different cell sizes are shown in \cref{fig:Resolution_SiCellSize}. The MLP method doesn't show a clear reference on the Si cell size. This conclusion may change with more sophisticated reconstruction techniques that exploit transverse granularity. At this stage, a Si cell size of \qty{5}{mm} is preferred, according to previous PFA-optimized designs.

\section{Conclusion and outlook}
\label{sec:conclusion}

In this study, several energy reconstruction methods were compared, including Sum E, N Hits, MLP, and DGCNN. In general, ML-based methods can significantly improve the energy resolution compared with traditional methods, with improvements of up to approximately 30\% at \qty{100}{MeV}. The MLP was chosen for the ECAL re-optimisation due to its higher computing efficiency and better performance at low energies. Conclusions for the re-optimized ECAL geometry can be drawn as follows:
\begin{enumerate}
\item With the energy leakage corrected by MLP, the ECAL material could be reduced to as low as \qty{18}{X_0} for \qty{60}{GeV} photons while maintaining the same performance, saving approximately one third of the total ECAL cost comparing to CEPC or ILD design\cite{group_cepc_2025,behnke_international_2013}.
\item Increasing the number of sampling layers continuously improves performance, but the improvement starts to saturate above 40 layers; Therefore, 30 layers are recommended.
\item Thicker Si sensors are preferred, and \qty{0.75}{mm} Si ECAL layers are planned for the next generation.
\item With current energy reconstruction methods, smaller cell sizes do not improve resolution for Si sensors \qty{0.75}{mm}-thick; the optimal SiW-ECAL cell size remains \qty{5}{mm}. However, this conclusion may change as more advanced energy reconstruction methods are developed, which could exploit the additional information provided by smaller cell sizes.
\end{enumerate}

This optimized geometry may contribute to future HET factories such as FCC-ee, CEPC, ILC, CLIC and others. The developed ML methods can be applied to other calorimetry studies, including ECAL test beam data and energy reconstruction for high-granularity HCAL. Furthermore, the results on the dimensions of the Si cells indicate that thicker Si is beneficial, which could be a requirement on the hardware side. On the software side.

Despite these results, several aspects remain to be explored. The conclusions regarding Si cell sizes may change when more advanced ML models are employed. The uncertainty introduced by hyper-parameter tuning is difficult to quantify in this study; however, it is supposed to be one-sided, as tuning always improves performance rather than degrades it. Therefore, this source of uncertainty is not expected to significantly affect the conclusions. Furthermore, more specialized ECAL layer designs, such as employing thinner absorbers in the first 10 layers followed by 20 thicker layers, could be investigated to fully leverage the ML-based energy reconstruction approach.

\section*{Acknowledgements}

The authors acknowledge the support from the Agence Nationale de la Recherche (ANR) under contract ANR-23-CE31-0023. 
\section*{Open Access}
For the purpose of Open Access, a CC-BY public copyright licence has been applied by the authors to the present document and will be applied to all subsequent versions up to the Author Accepted Manuscript arising from this submission.
\bibliography{SiW-Energy-resolution}

\end{document}